# How Push-To-Talk Makes Talk Less Pushy

Allison Woodruff and Paul M. Aoki
Palo Alto Research Center
3333 Coyote Hill Road
Palo Alto, CA 94304  USA
woodruff@acm.org, aoki@acm.org

## ABSTRACT
This paper presents an exploratory study of college-age students using two-way, push-to-talk cellular radios. We describe the observed and reported use of cellular radio by the participants. We discuss how the half-duplex, lightweight cellular radio communication was associated with reduced interactional commitment, which meant the cellular radios could be used for a wide range of conversation styles. One such style, intermittent conversation, is characterized by response delays. Intermittent conversation is surprising in an audio medium, since it is typically associated with textual media such as instant messaging. We present design implications of our findings.

## Categories and Subject Descriptors
H.4.3 [**Communications Applications**].

## 1. INTRODUCTION
Wide-area two-way radio service is increasingly available and popular in the U.S. Prior to 1996, only licensed radio operators were permitted to operate two-way radios that could communicate more than a few kilometers. In 1996, the deployment of digital cellular trunked-radio networks enabled wireless carriers to provide wide-area radio services to consumers. Just as in mobile telephony, subscribers can communicate privately with other subscribers (as opposed to using a shared public channel) at distances that are limited only by the network's cellular coverage; unlike mobile telephony, the service connects subscribers directly, without dialing delay.

One wireless carrier, Nextel Communications, provides mobile phones with conventional features such as voice telephony and voicemail; the same network and handsets also support a two-way, push-to-talk service called Direct Connect™. This service is very popular, having 10 million subscribers and supporting nearly 50 billion Direct Connect™ calls in 2001, predominantly for business use [16]. Competitors are attempting to introduce similar services based on packet (IP) networking; the top four U.S. carriers have all announced plans for similar services in the very near future, and separate service providers such as fastmobile (www.fastmobile.com) are also appearing, particularly in Europe.

While there have been a few in-depth studies of the use of short-range handheld radios, and the literature on the use of mobile telephony including text messaging (SMS) has expanded greatly in the last few years, we are not aware of any published studies of use of two-way, push-to-talk cellular radio systems (henceforth *cellular radios*). However, they deserve separate study because no other consumer service provides wide-area, private voice communication with a comparably lightweight interface.

We report here on a qualitative study of the use of cellular radios by consumers. The study, intended to inform the design of a new mobile voice communication system, was conducted as lightweight, exploratory "fieldwork for design." Our new system (described in part elsewhere [2]) has the design goal of supporting out-of-workplace social relationships within gelled social groups, especially those comprised of young adults. This population is interesting to study because its members allocate a great deal of time to social communication and value it highly. Accordingly, the study was aimed at collecting insights into emergent communication patterns developed by members of this target user population. At present, this population very rarely has access to cellular radios due to their cost. Therefore, we provided college-age students with cellular radios, observing their use of the devices and conducting interviews on an ongoing basis.

In our study, we observed a number of phenomena that can be associated with reduced interactional commitment. For example, participants did not feel they needed to reply immediately when someone spoke to them via the cellular radio (contrast this with a telephone conversation, in which people generally feel they must respond promptly when someone speaks to them). We further observed that these phenomena impacted the range of conversation styles available to the participants: the cellular radios supported a wide range of conversation styles – a range similar to that of instant messaging (IM) and wider than that of other audio technologies. As a result of their flexibility, the cellular radios were used in many different situations for many different activities. They were generally used in preference to other technologies, and participants reported they routinely used cellular radios when they would not have used other technologies.

The remainder of the paper is organized as follows. We first provide background and discuss related work. We then provide describe the method used in the study, including details of the cellular radio service itself. Next, we describe specifics of the participants' use of cellular radios during the study. We then discuss our findings, focusing on the different conversation styles that emerged. We then present design implications. Finally, we present conclusions and discuss future work.

## 2. BACKGROUND

In this section, we review a number of concepts from the literature that bear upon our observations and analysis.

*Conversation styles.* Research in mediated communication has made clear that a range of stereotypical conversational structures, which we will hereafter call "styles," occur in communication media. These conversation styles (Figure 1), which we discuss in turn below, occur to varying degrees in different media.

Much remote communication occurs through the telephone, using full-duplex audio connections (Figure 2, left)[1] that are set up on demand. Telephone interactions generally employ what we will call a *focused conversation* style in which participants formally engage in interaction (the *opening*), go through an additional period of conversational *turn-taking*, and formally dis-engage from interaction (the *closing*) (Figure 1, top) [8,20]. Within a sequence of turns, *lapses* in talk are highly noticeable – failure to take one's turn within a very brief time window is usually significant, indicating strong emotion, thoughtfulness, etc.

A different conversation style has been reported in studies of "lightweight" communication systems designed to facilitate "informal workplace communication" – the kind of "opportunistic, brief, context-rich and dyadic" [15] interactions that happen between physically proximate workers [26]. Such systems include classic open-channel environments such as full-duplex video spaces (e.g., [5]) and audio spaces (e.g., [1]), though there have been similar studies of messaging tools such as IM (e.g., [7,9,11,15]) that provide open-channel textual connections. A common finding is that the continuous availability of an open channel facilitates a *bursty conversation* style in which formal conversational openings and closings are infrequent even when long lapses occur between natural sequences of turns at talk (Figure 1, middle); the open channel puts the participants into what Schegloff and Sacks referred to as a "continuing state of incipient talk" [20], largely obviating the need for telephone-style openings and closings.

Studies of IM use have also highlighted communication behaviors that are now strongly associated with IM. Notably, Nardi *et al.* observed an *intermittent conversation* style in which long lapses can occur within what would normally form a single sequence of turns at talk [20], e.g., within a simple question-answer pair (Figure 1, bottom) [15]. Turn recipients who practice such lapses often draw upon *plausible deniability*, i.e., a reliance on the sender's lack of information (e.g., about the presence of the recipient at their computer), to excuse a lack of responsiveness. Lapses are also facilitated by the persistent nature of IM, since messages may be reviewed at a later time to recapture conversational context. "Hanging out" – the use of IM as a social space, with a number of IM sessions open and intermittently active – features prominently in teen and young adult use [7,11].

---

[1] In the telecommunications industry, *full-duplex* means that each participant in a channel can speak and hear others (send and receive) simultaneously, and *half-duplex* means that at most one participant can speak (send) at a time. Figure 2 shows an example of each. (Confusingly, half-duplex has a very different meaning in the radio operator community; for consistency, we adopt the telecom usage. Also, while it is physically possible for multiple radios to transmit at the same time, multiple transmissions interfere with intelligibility so severely that a single radio channel is half-duplex in practice.)

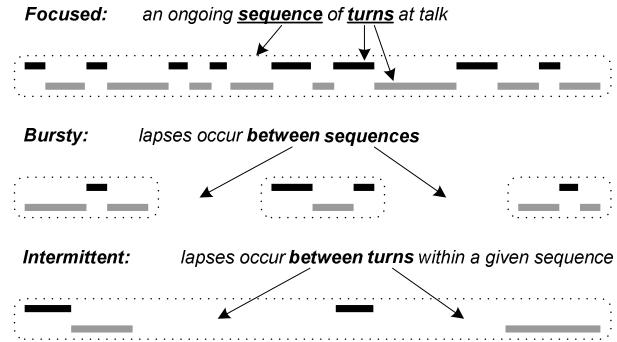

**Figure 1. Conversation styles.**

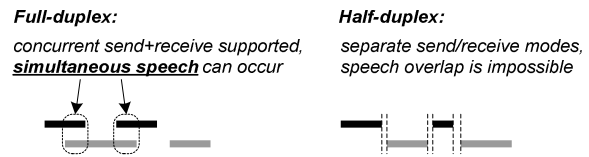

**Figure 2. Full-duplex vs. half-duplex channels.**

Nardi *et al.* also introduced the term *media-switching* [15]. Users have often articulated that media vary in their suitability for different purposes (e.g., IM is well-suited for brief messages and notifications whereas audio is good for communicating emotion through vocal affect [11]). Researchers have observed users beginning an interaction in one medium (e.g., a discussion in IM) and restarting it using a different medium that better suits their purpose (e.g., making a telephone call to repair a misunderstanding that occurred in IM) [9,15]. Apparently more common is the use of one medium (e.g., IM or SMS) to set up a communication in another (e.g., the telephone) [6,9].

Nardi *et al.* further discuss the notion that conversational participants negotiate an *attentional contract* that establishes that engaged communication can proceed [15]. (This is related to – in a sense, an inversion of – the "caller hegemony" inherent in telephone calls [8].) In this paper, we discuss the potential level of commitment that participants may make in such a contract.

*Mobile communication.* A recent research thread concerns social use of mobile communication media, including SMS [6]. Mobile phones, like IM, enable the construction of social spaces [10,12,23] though with the burden that such uses must be carefully managed in public spaces [18,24]. Mobile phones are widely used for *micro-coordination* – the use of dynamic, just-in-time activity coordination in lieu of extensive pre-planning [12].

*Two-way radio.* Two-way radio communication differs from most other audio media in that at most one speaker can be understood at a time (Figure 2, right). Radios are commonly used for wide-area communication but generally within specialist communities, which tends to limit the applicability of radio use studies to our work. In domains such as safety coordination, radio operators focus on minimizing miscommunication through formal procedures and standardized vocabularies. In (slightly) less formal domains, detailed studies have been made of the specialized communication and coordination work practices in railway control stations [13] and in amateur radio (e.g., [3]).

Studies of informal radio communication are of highest relevance to our work, since such communication involves use for casual purposes by untrained users. While use of informal portable radio (e.g., "walkie-talkie") communication is commonly reported, we are aware of little research that focuses on concerns relevant to ours. For example, Orr's study of handheld radio use by service technicians [17] emphasizes workplace adoption issues. Of particular note, however, is a study at Interval Research in which teens were supplied with portable radios during a weekend-long rock concert [22]; because of the choice of participants and "task" (i.e., socializing, partially structured around a common activity), this study found a number of novel behaviors and use-patterns.

To our knowledge, there have been no published studies of cellular radio, or indeed any other mobile technology that affords wide-area, half-duplex communication to the general public.

## 3. METHOD

In this section, we describe the procedure, participants, and equipment used.

### 3.1 Procedure

The study took place in June 2002. Participants completed a pre-study questionnaire on demographics and on their use of communication technology. Participants were given cellular radios and asked to use them for approximately one week and provide feedback. They were given almost no explicit direction; specifically, they were rarely if ever encouraged to use the devices *per se*, and they were aware that the study was being conducted by employees of a company that neither manufactured cellular radios nor provided cellular radio service.

One author lived with four of the participants in their rental house during the majority of the study, and participated in many social activities during the week. This participating author also observed several of the participants at work, and conducted semi-structured interviews before, during, and after the study. The participating author was able to observe transmissions made and received (on speaker) by co-present parties, as well as transmissions made and received by herself. The participating author took notes and recorded audio frequently during the study (approximately 50 hours of audio data was collected). After the study, the audio was reviewed and further notes and partial transcripts were made; an affinity clustering was performed on the resulting corpus of over 70,000 words.

### 3.2 Participants

We had two primary goals relating to the participants. First, we wanted the participants to be members of a pre-existing, gelled social group. Second, we wanted the participating author to be able to observe the participants throughout the day and night, in both public and private settings. We were able to accomplish both of these goals by working with a relative of the participating author. This relative was a participant in the study, and she selected and recruited all other participants by identifying the members of her own gelled social group. The participants accepted the participating author and gave her privileged access to intimate details of their lives, because her relative was a trusted member of the social group and vouched for her.

Participants were seven U.S. college undergraduates (five female and two male), all of whom were either 20 or 21 years of age and living away from their parents' homes. Most participants had known each other for several years and socialized frequently. Four of the participants (Erica, Julie, Ryan, and Todd) lived together in a rented house. Two additional participants (Kelly and Stephanie) rented an apartment together, and Dawn was a frequent visitor to that household. Julie and Todd were girlfriend and boyfriend. In addition to social and leisure activities, all participants were enrolled in summer school, working, or both. Of relevance to future discussions is that several participants worked as waitresses and Todd had a computer science internship. (Names have been changed to preserve participants' anonymity.)

In the pre-study questionnaire, participants reported that they very frequently used mobile phones: four of the participants had their own mobile phones, Julie and Todd shared a mobile phone, and Kelly did not have a mobile phone. Several participants frequently used IM, although others did not, e.g., one did not have a computer at home. Reported pre-study use of SMS, email, and home phones varied somewhat but was quite limited (this reported use was consistent with observed use during the study).

### 3.3 Equipment and system operation

We rented Motorola i1000 phones from a cellular service reseller. The phones were fairly large, measuring 114mm x 56mm x 30mm (4.5" x 2.2" x 1.2") and weighing 170g (6 oz.). Each participant, including the participating author, was given a phone and a single-earphone headset with a boom microphone; the phones could operate as a speakerphone (like a conventional handheld radio), as a telephone handset, or using a headset. To control costs (which are extremely high for voice calls on rental phones), all features except cellular radio service were disabled, e.g., the phones could not be used to place telephone calls. No limits, time or otherwise, were placed on the use of the cellular radio service. (The participants were free to carry and/or use any other communication technologies they wished.)

With this particular cellular radio service, connections between individuals are called *private connections*. One user uses a push-to-talk protocol to speak to another user. Specifically, if person A wishes to say something to person B, person A picks up their radio, selects person B's name from a "phonebook" menu, and holds down a button. After a brief delay (variable, but generally under 1 sec.), a "go ahead" beep is heard and person A can speak. (The radio uses short beeps to notify users of events such as the acquisition of the channel and the end of transmissions.) Person A's utterance is heard by person B as it is produced, with a network delay identical to that of a mobile phone call. When person A releases the button, transmission stops. After this, persons A and B simply push the button on their respective radios to speak (i.e., the menu is not used to select an addressee). Unlike conventional radios, the cellular radio system ensures that only one user can speak at a time using technological means – if person B pushes their button while person A is transmitting, person B's microphone will not be activated and the "go ahead" beep will not be heard until the channel is clear. After eight seconds have passed without a speaking turn, the radios reset and the phonebook menu can be used to select a new addressee; as an optimization, a "previous call" button reselects the last person whose transmission was received.

A similar *group connection* mechanism can be used within pre-specified groups of cellular radio users. This mechanism is very similar to a shared channel in a handheld radio, except that the network only allows access to group members. It is of limited usefulness, since users can only "tune in" to one group at a time, and, as with radio, they must already be "tuned in" to a given group to receive any of its messages.

A number of key factors distinguish push-to-talk interaction from telephone interaction. First, at an interactional level, the cellular radio has a lightweight model that does not involve an explicit act of call acceptance prior to answering. Second, at a practical level, a call between cellular telephones – even with "speed-dial" – takes many seconds of setup (dialing; up to a 6 sec. delay until the phone being called polls the network for incoming connections; ringing/pickup). By contrast, the cellular radio service advertises a mean setup time of 750 ms. This produces a qualitative difference in terms of spontaneity, i.e., between the initial "urge to speak" and the beginning of the first utterance. Third, a full-duplex telephone call affords inter-turn delays similar to those of typical face-to-face conversation; speakers can overlap, and often do so. By contrast, the half-duplex cellular radio channel inherently prevents overlap and paces conversation at a rate slower than face-to-face.

## 4. OBSERVED AND REPORTED USE

In this section, we discuss observed and reported use of the cellular radios. These findings are of course specific to the participants in this short-term study. However, we note that these findings are consistent with longer-term studies we have subsequently conducted.

Our goals in this section are to provide a general picture of the use of this technology by a gelled social group and to present the wide range of activities and communication patterns that emerged. Naturally, our investigation revealed many of the concerns and issues seen in previous studies of communication technology, e.g., privacy and availability. However, due to space constraints, we are unable to discuss all of these points at length.

### 4.1 General patterns

In this subsection, we review the general use of the cellular radios.

*Overall availability.* Participants typically carried their cellular radios with them and kept them turned on. For example:

> Dawn: "I learned that I should just take it with me into the shower in case somebody's like trying to talk to me."

Participants generally left their radios on at night and were sometimes awakened by them. (Interestingly, the ritual "good morning" message [1,15] and "goodnight" message [6,23] usually observed with lightweight communication technologies did not come up in observation or in interviews.)

*Overall level of activity.* While most participants were very strong adopters (we roughly estimate that participants used their cellular radios for on the order of tens of interactions per day), the level of activity was extremely variable. Sometimes there would be great bursts of activity, which were often interleaved with other social activity. For example, at a given moment in a car with five people, two participants in the car might be using their cellular radios in private connections with participants at other locations, while simultaneously participating in co-present conversations with people in the car. Such activity was often chaotic, especially since the cellular radios were typically used in speaker mode and many conversations involved a lot of vocal affect. At other times, participants would not use the cellular radios for long stretches, e.g., while they were engaged in an activity like watching a show.

*Dominant modes of use.* Participants used private connections much more frequently than group connections, partly because the group mechanism was cumbersome and partly to avoid annoying other participants. Dyadic connections occurred among most pairs of participants, although some dyads conversed more frequently than others.

Participants explained that speaker mode was preferable to headsets because the headsets were uncomfortable and because they drew attention from other people. Early in the week, the visual appearance of the headsets with the boom microphones was a source of amusement: several girls joked extensively that they were a girl band, and one boy dressed up in a security guard shirt that he happened to own. This type of affiliation has been reported elsewhere; recall that in [22], teens using short-range radios pretended to work for concert security.

Because the devices were used primarily in speaker mode, co-present individuals were able to overhear transmissions and consequently frequently became involved in interactions. Similar impact on co-present individuals has been observed for other media as well [5,18,24]. Neither transmitters nor receivers seemed particularly sensitive to the public nature of transmissions, although they did indicate some embarrassment about the fact that speech emanated from unexpected areas of their body, depending on where they had clipped their cellular radio. This sense of body parts such as thighs or hips "talking" did not appear to result in a change in practice, i.e., they continued to use speaker audio and clip the radios to the same areas.

### 4.2 Comparison to use of other media

Participants said they used the cellular radios in preference to other technologies. In some cases, they used multiple communication media simultaneously; in these situations, the cellular radio had lower priority than media such as the telephone.

*Frequency of use.* Participants reported, and we observed, that they spoke much more frequently on the cellular radios than on their mobile phones. Cost did not seem to be a significant factor in the choice of medium, particularly as the participants generally had the cellular phone contract plans that are common in the U.S. (as opposed to the pay-as-you-go plans that are more common in Europe). Instead, participants reported that they were strongly influenced by the fact that the communication was lightweight in comparison with mobile phones. These findings are consistent with subsequent studies in which we provided unlimited cellular radio and unlimited cellular phone service in the same device.

> Dawn: "I didn't have to do any button, dialing thing, plus I don't remember her number."

> Erica: "[I]t seems like so much work to call [telephone] them."

Participants identified many specific situations in which they would use cellular radios in which they would not use mobile phones. For example, Erica reported that she would not use the phone to chat with people while she was at work, but she would use the cellular radio because she could start and stop the conversation quickly.

> Erica: "[I]f it weren't for the walkies talkies I just wouldn't talk to them."

Similarly, participants reported that they used the cellular radios in many situations in which they did not believe they would have used SMS (although most had fairly limited experience with it), saying that SMS was undesirable because of the effort of typing, the long delays, and the difficulty of communicating affect in text.

People sometimes relayed or requested information over the cellular radios that would probably not have been worth sharing using more heavyweight mechanisms (analogous to effects reported for SMS relative to telephony [10]). For example, Erica said she would call people to ask questions which she felt would not be appropriate with the phone.

> Erica: "It's really convenient with roommates. Cause you can ask em just stupid little questions, like, you know, 'Where's the extra toilet paper?' or something."

*Media use within-group vs. outside-of-group.* Cellular radios were almost always chosen in preference to mobile phones for communication within the group of study participants (note that participants continued to use their mobile phones for communication with people outside the group).

> Dawn: "[T]hat would be like the drastic emergency thing. If I couldn't get anybody through this [the cellular radio], I would have to be like, okay, I have to use the phone now."

For at least one participant, cellular radio replaced the use of IM with other participants. Note that media-switching from cellular radios to other technologies was not observed or reported.

Cellular radios did not appear to have much impact on the use of communication technology or the frequency of communication with people who were not in the study. In direct terms, this is in part to be expected since almost none of the people the participants knew outside of the study owned cellular radios. However, in more indirect terms, one might expect a kind of "conservation of talk" – that intensified within-group communication might result in reduced communication with people outside of the group. This did not seem to occur.

*Precedence in cases of conflict.* Cellular radios were sometimes used at the same time as other communication media. Telephone calls, when they did occur, took precedence over cellular radios. For example, if a participant were on the telephone and received a transmission on the cellular radio, they would typically ignore it.

> Julie: "[A cellular radio transmission is] not like a phone call, so it's like a lower priority…"

Because of the prevalent use of speaker audio, it was not unusual for someone nearby to pick up the participant's cellular radio and reply on their behalf.

## 4.3 Communicative activities and purposes

The cellular radios were used for a wide range of activities and purposes, which exemplify all the different conversation styles discussed in the related work section. In fact, a given activity would often involve multiple communication styles, with the participants fluidly moving among the different styles without explicitly negotiating the change. For example, an extended remote presence interaction might transition from bursty conversation to intermittent conversation if a participant became busy and started delaying their responses.

While many of these activities have been reported for other media, the diversity of uses for a single audio medium is interesting. We emphasize that all these uses were originated by the participants.

*Chit-chat.* The cellular radios were frequently used for brief, light conversation, particularly by the females. This often occurred as a "time filler" when participants were bored. For example, the girls who worked as waitresses would often use the cellular radios to talk with other people when they were bored at work. Casual conversation was also a popular activity when people were walking or driving somewhere by themselves.

> Erica: "It's nice like walking back from work and stuff to be able to call people and just to chit-chat."

> Dawn: "Kelly called me one time from her car, she's like, 'There's a fine guy in front of me, I'm following him!' I'm like, 'Go for it!'"

*Extended remote presence.* The cellular radios were often used for sharing extended activities or to allow participants to keep each other company for extended periods. We distinguish this from remote single-task participation, such as consultation during a trip to the grocery store. The difference is that what we call extended remote presence would continue for a very long period of time relative to a plausible phone call length; the periods could safely extend across local activities in which one participant or the other would have been likely to hang up a phone call. Julie "went on errands" with Kelly by speaking to her periodically in the cellular radio. While Erica and her boyfriend were at a baseball game, they checked in from time to time with participants who were eating dinner at a restaurant. The girlfriend/boyfriend pair kept in contact, speaking often while Todd was at work.

> Julie: "I talk to Todd a lot to just, you know, see what he's doing at work or just to bug him."

Another variation was interaction during sequences of short tasks. For example, Julie reported that she taught Ryan to make rice. She said the cellular radios were convenient for this; there was a lot of waiting while he completed steps in the recipe, so a phone call would have been inconvenient.

*Micro-coordination.* Cellular radios were often used for micro-coordination of shared activities [12]. For example, one group went into a grocery store while another group went through a drive-through at a fast food restaurant; the group in the store spoke to the group in the drive-through about their order and then coordinated being picked up in the store parking lot.

*Substantive conversation.* The cellular radios were used for substantive conversations which were often multi-topical and sometimes lasted as long as thirty minutes. These were often very similar to interactions that might occur on the telephone. Such interactions often focused on the sharing of feelings and emotional support; for example, Erica contacted Kelly to tell her a long story about a guy who was a "jerk" at work. The ability to communicate affect through voice was important [11].

> Julie: "I like it [the cellular radio] better than IM cause like in IM you always lose everyone's… intonations when they say stuff…"

*Play.* The cellular radios were used for a number of playful activities. These largely relied on vocal "sound effects" (as in [1,22]) and the communication of affect, and so would not have been appropriate in a textual medium such as IM. They would also have been difficult to carry off if the participants had had to call each other on the phone. For example, the two boys would frequently play military games like pretending to land airplanes, or spontaneously say things like "Damage report in sector 4" to each other. Other play included spontaneous singing of songs or repeated quoting of phrases from movies in funny voices.

Additional jokes relied on the fact that the cellular radios were used primarily in speaker audio mode, which created a space that drew in co-present parties [5]. For example, Dawn spoke to Julie's pet hamster without preface, saying, "Butterscotch, hello" at a time when she guessed that Julie might be near her hamster. In fact, pets were addressed and/or encouraged to "speak" during transmissions on multiple occasions.

## 4.4 Reduced interactional commitment

Participants had a strong sense that contacting someone on the cellular radio did not represent a commitment to a full-fledged conversation. In contrast, when interactions take place in media such as the telephone, people are generally understood to have made a full commitment to participate in an interaction and to give it their exclusive attention. Participants considered the reduced commitment of cellular radios to be an advantage.

> Julie: "Like a phone call is a really big commitment for me. You know, it's like I, I totally plan phone calls… I don't call people to just say like, hey, what are you doing?"

In this subsection, we discuss several communication phenomena that occurred in the activities described above and exemplify a form of reduced commitment. We discuss how the half-duplex and lightweight aspects of the cellular radios were particularly relevant to these phenomena.

*Reduced openings and closings.* Openings and closings were generally omitted or reduced as compared with interactions in other media, such as telephone conversations. Participants did not feel that individual cellular radio interactions required many formalities. Such formalities could be considered the negotiation of an attentional contract, and one can argue that higher levels of commitment require more formal agreement. Apparently, the reduced commitments being made for cellular radio interaction required little formality.

> Dawn: "[Y]ou don't have to be proper, hello, blah blah blah blah blah, conversation, conversation."
>
> Erica: "This, there's no hanging up. It's just putting it away."
>
> Todd: "It's just kind of a more immediate channel… it's something that without much effort you can just kind of engage them… I mean, phones kind of have this thing, I don't know, it's kind of a, a commitment to call someone on the phone and then you have to, you know, have this conversation."

The lightweight nature of the cellular radios made it easy for participants to contact each other, and therefore encouraged reduced openings and closings by creating the illusion of a space. Both reduced commitment and reduced formality are typical of informal face-to-face communication [26] and are commonly reported in systems in which users remain continuously connected, such as media spaces (e.g., [1]) and IM (e.g., [15]).

*Delayed or omitted responses.* Participants often delayed their responses to cellular radio transmissions. The participating author observed delays of approximately two to three minutes that occurred without apology or explanation. At other times, transmissions would simply go unanswered.

> Kelly: "[I feel like I have to answer if somebody says something to me] but not immediately. I can do it on my own time… if I'm like busy or something like that, and then when I get a chance I'll be like, what did you say, what do you want?"

The participants broadly accepted the type of behavior Kelly describes, although their tolerance varied somewhat by context.

> Kelly: "[T]here can be long pauses and nobody cares and so, phones are just so restrictive and the fact that you have to pay attention so much."

Delays were often attributed to the likelihood that the non-responsive participant had become engaged in another activity, e.g., that their boss had come into their office.

> Erica: "I understand to wait if I'm talking to anybody till they're free and stuff [if they don't answer]… I tried to message Julie earlier but it wasn't working and I figured she was probably at work, busy in a meeting or something."

While we are not claiming that delayed or omitted responses were universally acceptable, cellular radios appeared to facilitate delayed or omitted responses because their half-duplex nature affords the kind of plausible deniability evident in IM [15]. Specifically, a sender has little or no information about the status of a recipient – the cellular radio itself provides no ongoing awareness information (the half-duplex nature of the channel means that information is only received when participants make explicit transmissions), and there is no IM-like supplementary presence mechanism. Hence, the recipient is less visible to the sender and less accountable to respond. Additionally, because the media is neither persistent nor entirely reliable, accountability is further reduced: there is no guarantee that messages are received.

Note that we have not conducted a comparative study of the occurrence of this phenomenon in IM and cellular radio, e.g., our data does not give us information about the relative frequency of delayed or omitted responses in the two media.

*Reduced feedback.* Because the channel is half-duplex, by definition only one participant could speak at a time. This made it difficult to give feedback (including the utterances that are often termed "backchannel" or "continuers"). Workarounds were generally unsatisfactory; several participants reported transmitting "fake" (non-spontaneous) laughter after another participant completed a funny utterance. While this inability to provide natural feedback in overlap with the other participant's utterance was occasionally frustrating, the participating author observed that it was somewhat liberating not to have to respond continuously, e.g., listeners were not under the same obligation to make sympathetic noises in response to a story as they would be in, for example, a telephone or face-to-face conversation.

*Interleaved activity.* Many activities were routinely interleaved with use of the cellular radios.

> Kelly: "On these things [the cellular radios], I can be typing da da da da da and listen to someone say something, they're not offended that I'm typing on the computer and then I can pick up and say de de de de…"
>
> Todd: "[T]he phone is more formal and it… takes all of your attention, and the radio can take all of your attention, but you can kind of also kind of back burner it a bit."

Half-duplex facilitated interleaved activity. Because the channel is one-way, information did not "leak through." In other words, when a participant was transmitting, they did not hear audio from the recipient of the transmission. For example, a transmitter could not hear that a remote recipient was typing, conversing with other co-present people, or otherwise failing to pay full attention.

## 4.5 Social impact

The cellular radios impacted the participants' social lives in a number of ways. Participants found that they spoke to each other more frequently and consequently had more awareness of each other's activities. The combination of more frequent conversation and increased awareness led to them seeing each other more often.

*Increased remote communication.* As discussed above,

participants said they used the cellular radios more frequently than other technologies; as a direct consequence, they spoke to each other more often. Further, because it was so easy for participants to contact each other, expectations increased.

> Todd: "You know if you have it, people expect to be able to talk to you all the time."

Feelings about this increased availability were mixed. Some were pleased.

> Julie: "This is so great. I just lie here and all these people talk to me."

Others were more ambivalent. For example, while some participants liked being able to reach other people easily, those same participants sometimes found it irritating that other people were able to reach them easily.

Overall, increased availability was tolerable largely because there was a limited group of participants using the cellular radios, and therefore they were available only to close friends. They contrasted this with mobile phones, which they felt gave more people access to them.

> Julie: "[I]t's kind of fun too, to have like this network where it's like only my friends and only like fun people are calling me. So I kind of like that."
> Kelly: "Yeah, I like that too."

When given hypothetical long-term use scenarios, participants were clear about the types of people with whom they would be willing to share cellular radio connectivity, and therefore increased availability. Participants were most interested in using cellular radios with roommates and close friends whom they saw regularly, they were less interested in using them with friends who lived far away, and they were generally strongly opposed to using with their families. Although the desire for young people to control or avoid contact with their families has been documented elsewhere [12], the threat of using cellular radios with family members appeared to be even more severe.

> Todd: "[I]f my dad had a radio-"
> Ryan: "Oh, my God."
> Todd: "I would just be in constant sorrow for all my days."

*Activity awareness.* Participants said they had more information about what the other participants were doing.

> Erica: "I know where everybody is. Like, I usually don't know where Kelly and Dawn are. Like I, I just don't keep up with them that close, to know what they're doing… now I know when people are working. So much better this week."

This was because of the nature of the communication as well as the increased frequency. For example, participants said they often used the cellular radios to ask each other "What's up?" or "How's work?" Participants were positive about this increased awareness.

> Julie: "I actually know what Todd's doing at work today. Cause usually [it's] like 'How was work?' 'Suck. Nothing, nothing happened. I'm so bored.' This is much better."

However, we did not get the sense that the participants had the high degree of awareness that has been documented for other systems such as video spaces [5].

*Increased co-present social interaction with participants.* The participants felt that they saw each other more frequently while they had the cellular radios, and they were pleased by this.

> Julie: "I think like I never see Kelly this much. Ever. Like sometimes I get to see her on the weekend."

While they felt the amount they saw each other was affected somewhat by activities that were taking place that particular week, e.g., some of the girls were leasing a new apartment, they clearly articulated that both more frequent talk and increased awareness were key factors in the increased visitations. More frequent talk provided more openings to coordinate co-present activities, as well as being a resource for learning that initiating such activities would be appropriate.

> Todd: "[T]he opportunity to talk with someone more usually leads to like, you know, do you want to go do blah blah blah. And. That just seems kind of a natural thing for me."

## 4.6 Overall subjective response

Participants said they liked the cellular radios and that they were fun. When asked if they would want to have the service long-term, they were moderately (but certainly not overwhelmingly) enthusiastic. After the study, some of the participants missed their cellular radios and said they were bored without them.

> Julie: "I wanted to talk – I was so bored on my walk home. From class. 'Cause I had no one to talk to. It was really sad."

## 5. DISCUSSION

In very broad terms, the consensus picture that emerges from recent research on personal communication (Section 2) is that youth in much of the industrialized "First World" communicate using a variety of media. These users demonstrably employ each of the conversation styles we have described (focused, bursty, and intermittent) and select communication media as necessary to address their current needs. Factors in their selection include the characteristics of the medium itself, but this is one of many factors (including economic cost, purpose, and physical environment).

Given this context, it is striking to recall that participants used their cellular radios for nearly all mediated communications within their social group (Section 4.2), using them in a rich variety of communicative activities that demonstrated the entire range of conversation styles (Section 4.3). Such a wide range of conversation styles is unusual in a single medium, and to our knowledge, the range of conversation styles we observed with the cellular radios has not been reported previously for any audio medium. It would be difficult to tease apart all of the factors that drove the participants' media selection. However, as designers, we wanted to understand how the participants had been able to appropriate the cellular radios for such a wide range of conversation styles. That is, we wanted to have some idea about which aspects of the technology were critical to this flexibility so that we might afford this same flexibility in designs for future communication systems.

In this section, we describe aspects of the cellular radios relating to their use for different conversation styles. We then discuss how this range of styles impacted the participants.

## 5.1 Diversity of conversation styles

In our discussion of how features of the cellular radio system affected users' ability to apply different conversation styles, we draw upon a comparative analysis of multiple communication media, including cellular radio. The analysis (discussed more fully in a technical report [27]) applies elements of mediated communication theory, which explores "the relationship between the affordances…of different mediated technologies and the communication that results from using those technologies" [25]. In particular, we consider the cellular radios in light of Clark and Brennan's theory of communicative grounding [4], which goes beyond simple notions of media "information richness." A variety of specific features of the cellular radios create a sense of reduced interactional commitment, resulting in the reduced commitment phenomena discussed in Section 4.4 (i.e., reduced openings and closings, delayed or omitted responses, reduced feedback, and interleaved activity) and, in turn, the different conversation styles. As we will see, the most significant features relate to the half-duplex and lightweight aspects of the cellular radios.

In this subsection, we discuss each of the conversation styles in turn. For each conversation style, we discuss whether it might be expected in cellular radio use. Then we show how this conversation style was facilitated by the cellular radios, particularly the reduced commitment phenomena. (For each conversation style, we discuss only the most relevant phenomena.)

*Focused conversation.* Focused conversation is characterized by highly attentive and responsive interaction (Figure 1, top). While focused conversation is common in other audio media, such as telephony, it is somewhat surprising to see it occur with cellular radios. Anyone who has used a walkie-talkie for extended periods can attest that sustained, turn-by-turn conversation takes significant effort, and some participants did state that the push-to-talk nature of cellular radios made substantive conversation more difficult. The issue was not simply the slower pace, or the fact that the device required participants to press a button to speak; the one-at-a-time nature of the channel precluded participants from giving feedback while someone else was speaking. That is, while feedback was possible, it could not be positioned in what would be the natural place (i.e., in overlap) relative to the other participant's speech in a face-to-face or telephone conversation. In general, we observed that participants produced less feedback, which is known to reduce the fluidity of turn-taking [25].

Overall, however, participants expressed surprisingly little concern about the half-duplex channel as a major inhibitor to substantive conversation. In fact, the cellular radios were largely preferred for this purpose to other technologies that enabled more fluid turn-taking, such as mobile phones (recall that media-switching from the cellular radios to phones was not observed or reported). At least two phenomena contributed to this preference. First, the reduced feedback offered a subtle advantage for the focused conversation style: since a listener was not required (or indeed able) to give any feedback while someone else was speaking, the half-duplex conversation reduced the level of effort required to participate in the conversation. Second, participants highly valued the sense that reduced closings were acceptable, which meant that interactions could be ended quickly. Becoming mired in an undesirably long conversation is often cited as a reason to avoid making phone calls (see, e.g., [6]), and the ability to close quickly was valuable, e.g., when one was called on to wait on a table.

*Bursty conversation.* Bursty conversation is characterized by multiple brief, focused sequences of turns at talk with reduced openings and closings (Figure 1, middle). If one considers the most obviously similar communication media, one might not expect to see bursty conversation in cellular radio. Bursty conversation is not generally reported in telephone use, as calls are generally closed when a new sequence does not arise. Amateur radio, which seems even more similar because it too involves half-duplex audio, uses conversational protocols [3] that formalize (as opposed to reducing) openings and closings.

In practice, the cellular radios appeared to be more similar to media spaces, in which bursty conversation is known to be common. The lightweight nature of the cellular radios encouraged reduced openings and closings. Further, the cellular radios could be used for focused turn sequences, as discussed above. As the cellular radios enabled both of the key elements of bursty conversation (reduced openings and closings, and focused sequences of turns), participants could and did use them in this manner quite frequently.

*Intermittent conversation.* Intermittent conversation is characterized by lapses in talk between individual turns. Unlike bursty conversation, the current sequence of turns may not be appear to be anywhere near completion – a participant who might be expected to respond, e.g., to a question, simply does not do so for an extended period of time (Figure 1, bottom). The literature associates this style almost exclusively with IM, and IM is very different from cellular radio in many dimensions. It is certainly the case that we do not know of any other audible media in which intermittent conversation has been reported, and in fact the immediate nature of real-time voice communication often makes recipients feel compelled to answer immediately when addressed. For example, users of the Thunderwire audio space who were known to be at their desks (e.g., those who had just been heard in the space) would signal inattention explicitly if they could not answer promptly [1]. Further, easy reviewability of messages is often cited as an important enabler for many kinds of temporally decoupled interactions in IM [9,15]. The non-persistent nature of cellular radios means that inattention risks loss of information – the pacing of voice communication requires recipients to hear and mentally buffer the utterances as they arrive.

Given this context, it was surprising to find that participants felt that intermittent conversation was a core use of the cellular radios. Participants saw the cellular radios, although audible, as compatible with intermittent conversation, and independently volunteered their perception that the cellular radios were more closely related to IM than to the telephone.

> Kelly: "I think it's really close to IM. Like I really like it that it's so close because you know, it's one message at a time, it's you know, not commit like, not, you don't have to talk for a long time, you can like leave if you want to or like not answer… [It's more like] IM than the phone."
>
> Erica: "It's kind of like IM over the phone..."

As discussed above, delayed or omitted responses were considered acceptable in part because the half-duplex channel provided little information about the participant's availability. Further, interleaved activity was facilitated by the half-duplex nature of the interaction, which concealed the sound of other activities such as typing so such activities did not "leak through" and offend other participants by implicitly signaling inattention. (Contrast this with most audible media such as telephones and audio spaces which are full-duplex so such activities can be heard by the other participant(s).) Half-duplex communication, which is usually considered to have negative effects on spoken conversation [25], may in this case be a key enabler for IM-like interaction styles in audio.

## 5.2 Impact of different conversation styles

All three conversation styles are useful in particular situations, as illustrated by our study as well as other studies reported in the literature. Unlike most other media, cellular radios adequately supported all conversation styles – focused, bursty, and intermittent conversation each occurred frequently and spontaneously in the participants' cellular radio use. Changes between different conversation styles occurred fluidly, without explicit negotiation. Participants found the cellular radios to be adaptable enough to meet most of their mediated communication needs, and media-switching apparently became unnecessary.

However, while the cellular radios appear to have been *sufficient* for most needs, they were not necessarily *optimal* in all situations. For example, lengthy conversations often occurred but required quite a bit of work in the half-duplex channel.

> Erica: "[I]f you need to have an actual long conversation, telephone [is preferable]… It's kind of annoying when you're trying to have a conversation with somebody, you have to wait till you know they've finished their thought, wait a couple seconds, press the button, wait a second, and then talk, finish your thought…"

While Erica expressed that the telephone was preferable for focused conversation, in fact she used the cellular radio for such conversations (see, e.g., Section 4.3). Similarly, Julie explained that she considered switching to the telephone but did not:

> Julie: "I was wondering… while we were doing it [having a long conversation]… I'm wondering if I should call her…"

Recall that participants were reluctant to commit to telephone conversations; this may have been a factor in the continued use of the cellular radios for focused conversation. It is possible that if the participants had known in advance that they were going to speak for fifteen minutes, they would have used the telephone instead of the cellular radio. (In practice, of course, they were unlikely to have that knowledge, since neither party had perfect information about the other's availability and environment.)

Participants generally did not switch to other media as conversations evolved, even when it appears a switch may have been beneficial. People may in fact generally be reluctant to media-switch (e.g., consider the low rate of media-switching observed by Isaacs *et al.* in IM [9]). In the next section, we discuss some applications of these findings to design.

## 6. FROM FIELDWORK TO DESIGN

Designers continue to create novel voice communications systems for mobile and ubiquitous computing environments (e.g., [2,14,21]), particularly as bandwidth to support continuous network connectivity becomes increasingly available. The flexibility of the cellular radios suggests a number of issues for designers of future voice communication systems. In this section, we first discuss implications of our findings for voice communications systems. We then describe a design concept for a novel adaptive channel system which we are currently building.

*Implications for voice communication systems.* While designers typically strive to increase the feature-richness and "media richness" of their systems, our findings (e.g., Section 5.1) suggest that features that reduce interactional commitment are desirable in many situations even though such features may be associated with limited functionality or "low quality" communication.

As a first example, consider that half-duplex audio is often considered to be inferior to full-duplex audio, causing disruption of normal conversational behavior (see, e.g., [25]). With the cellular radios, lightweight, half-duplex audio resulted in a useful balance between immediate access and relatively low interactional demands on the participants. This balance meant that the participants were willing to use the cellular radios during most of the day and night. In contrast, full-duplex audio demands higher degrees of engagement which may not be as tolerable in continuous long-term use (particularly in social, mobile conditions). Overall, half-duplex audio may therefore provide many of the benefits of continuous full-duplex audio spaces while ameliorating some of the key disadvantages (such as loss of privacy and overload) that are typically associated with them.

As a second example, consider the common addition of recording features to voice systems, such as the reviewable "audio chat" implemented by Impromptu [21]. While persistence features have advantages, our findings suggest that such features would disrupt the plausible deniability currently afforded by the cellular radios; the "burden" of reliable communication may increase the tendency of users to turn off the system rather than risk receiving "undeniable" messages at particular times. Similarly, persistence may work against spontaneity; our participants were willing to use a non-persistent medium to make trivial but affect-rich comments since the cost to the recipient was low, but may have been unwilling to make these same comments if they knew the comments persisted as entities that the recipient had to spend time to manage (like voicemail or email messages).

*Proposal for a new kind of adaptive channel.* Our findings (e.g., Section 5.2) suggest that it is highly desirable to support users in their moment-by-moment changes of conversation style with maximal fluidity, i.e., without requiring them to switch devices, change applications or even conduct an explicit negotiation. As discussed above, the act of explicit media-switching has interpersonal interaction issues which remain even if user interaction issues are minimized. A better formulation might focus on what we will call *style-switching*. Our study participants were able to do this (albeit perhaps suboptimally) by simply using the same medium in different ways. We suggest that technological means can be used to adapt a medium to participants' conversational needs, in a manner that goes beyond media-switching or multimodality.

As an example, consider a system that monitors participants in an ongoing conversation and automatically adapts *properties* of the channel – properties that have, in the past, been fixed for a given technology, such as half-duplex vs. full-duplex – in response to observed characteristics associated with different conversation styles. Such monitoring can be of the individual participants (e.g., their observable emotional state [19]), or of their interaction (e.g., their turn-taking engagement with other participants [2]). For example, imagine two participants in a push-to-talk session, each responding slowly (intermittent conversation) because they are both engaged in other tasks. Now suppose that a new topic of conversation is raised and both participants become highly interested. The system may detect that the participants are showing strong signs of interest (e.g., their voices have acoustic properties associated with interest) and that they are showing signs of increased conversational engagement (e.g., they begin to respond much more rapidly than before), concluding that a focused conversation has begun. In response, the system shifts the channel to an open-microphone, full-duplex mode, playing a short tone to indicate that push-to-talk will no longer required. Later, when the demonstrated level of engagement dies down (e.g., by a sustained pattern of lapses between turns), the system shifts the channel back to push-to-talk.

A key advantage of such a system over negotiated media-switching is that, by relying on (conservative) measures of engagement, it allows the interaction to proceed in a way that is in accordance with an implicit but demonstrable attentional contract while finessing an explicit but potentially disruptive negotiation.

## 7. CONCLUSIONS AND FUTURE WORK

We observed many communication phenomena associated with reduced commitment to the current conversation. Specifically, participants exhibited reduced openings and closings, reduced feedback, delayed or omitted responses, and interleaved activity. These phenomena appear to have been enabled or facilitated by the half-duplex, lightweight nature of cellular radio transmissions. This reduced commitment appears in turn to have facilitated a wide range of conversation styles: participants used cellular radios for focused conversation, bursty conversation, and intermittent conversation, fluidly moving among these different styles without explicit negotiation. This is an unusually broad range for a single medium, and to our knowledge intermittent conversation has not previously been reported for an audio medium. We are currently working on the design and development of an audio system that we hope will support this range of styles even more effectively.

In this work, we used cellular radio service as a rough approximation of our own lightweight audio system to identify emergent issues and phenomena for a specific population. One could obviously extend the study in a number of ways, e.g., increased numbers of participants, longer term of use, etc. Additionally, it would be valuable to directly compare the phenomena for different media. For example, are delayed responses less acceptable with cellular radios than with IM?

## 8. ACKNOWLEDGMENTS

We are grateful to Beki Grinter, Peggy Szymanski, and Jim Thornton for many helpful discussions.